\documentclass[12pt]{iopart}

\usepackage{iopams}  
\usepackage{graphicx}
\usepackage{siunitx}
\usepackage{booktabs}
\usepackage{xcolor}
\usepackage{cite}  
\usepackage[hidelinks]{hyperref}

\usepackage[explicit]{titlesec}
\usepackage{ulem}
\usepackage{lipsum}

\titleformat{\paragraph}[runin]
  {\normalfont\normalsize\bfseries}{}{15pt}{{\theparagraph\hspace*{1em}#1.}}
\titleformat{name=\paragraph,numberless}[runin]
  {\normalfont\normalsize\bfseries}{}{15pt}{{#1.}}

\begin{document}

\title{Hadrons, Better, Faster, Stronger}

\author{Erik Buhmann, Sascha Diefenbacher, Daniel Hundhausen, Gregor Kasieczka \& William Korcari}
\address{Institut f\"ur Experimentalphysik, Universit\"at Hamburg, Luruper Chaussee 149, 22761 Hamburg, Germany}
\ead{sascha.daniel.diefenbacher@uni-hamburg.de}

\vspace{10pt}

\author{Engin Eren, Frank Gaede, Katja Kr\"uger, Peter McKeown}
\address{Deutsches Elektronen-Synchrotron DESY, Notkestr. 85, 22607 Hamburg, Germany}
\ead{engin.eren@desy.de}
\vspace{10pt}

\author{Lennart Rustige}
\address{Center for Data and Computing in Natural Sciences and Deutsches Elektronen-Synchrotron DESY, Notkestr. 85, 22607 Hamburg, Germany}

\vspace{10pt}

\begin{indented}
\item[]December 2021
\end{indented}

\begin{abstract}
Motivated 
by the computational limitations of simulating interactions of particles in highly-granular detectors, there exists a concerted effort to build fast and exact machine-learning-based shower simulators.
This work reports progress on two important fronts.
First, the previously investigated WGAN and BIB-AE generative models are improved and  successful learning of hadronic showers initiated by charged pions in a segment of the hadronic calorimeter of the International Large Detector (ILD) is demonstrated for the first time.
Second, we consider how state-of-the-art reconstruction software applied to generated shower energies affects the obtainable energy response and resolution.
While many challenges remain, these results constitute an important milestone  
in using generative models in a realistic setting.
\end{abstract}


%
%
%
%
%

\newpage
\section{Introduction}
\label{sec:intro}

Precise simulations of interactions between fundamental particles and complex detectors are a prerequisite for carrying out modern particle physics research. 
The high computational cost of these simulations is well established~\cite{HEPSoftwareFoundation:2017ggl}
and has --- sparked by Ref.~\cite{CaloGAN2}  --- resulted in a large scale effort to develop surrogate simulations based on generative machine learning models.
Trained on a small initial dataset --- produced either using classical, Monte-Carlo-based simulators such as \textsc{Geant4}~\cite{g4} 
or potentially taken from  data ---
these generators amplify the effectively available statistics~\cite{Butter:2020qhk}.

Motivated by the large fraction of resources already consumed by calorimeter simulation~\cite{atlas_fastsim}, and the expected increase due to higher granularities and luminosities, the precise and fast simulation of calorimeters is a primary topic of research~\cite{CaloGAN2,CaloGAN,CaloGAN3,ErdmannWGAN,ErdmannWGAN2,Sofia,ATLAS_Gen,ATLAS_Gen2, ATLAS_Gen3,Buhmann:2020pmy,decoding_photons,Khattak:2021ndw,Carminati:2020kym,Hariri:2021clz,Rehm:2021zow,Rehm:2021zoz,Rehm:2021qwm,Krause:2021ilc,Krause:2021wez}. \footnote{For reviews, including also other applications of generative models to particle physics, see Refs.~\cite{Alanazi:2021grv,Butter:2020tvl}.}

In the quest to simulate calorimeters with high accuracy, standard generative machine leanring methods --- Generative Adversarial Networks (GANs)~\cite{GAN}, Variational Autoencoders (VAEs)~\cite{VAE} and normalising flows~\cite{NICE,RealNVP,rezende2015variational,papamakarios2019normalizing} --- have all been considered.
However, several obstacles need to be overcome before these tools can be deployed to simulate highly granular calorimeters with high resolution.
In Refs.~\cite{Buhmann:2020pmy,decoding_photons} we showed a precise modelling of differential distributions over many orders of magnitude for electromagnetic showers. The present work extends this level of precision for the first time
to the more challenging hadron-induced showers in a highly granular hadronic calorimeter.

Another important aspect is the downstream processing of generated showers. 
While so far the focus has been on the \textit{raw} output of the generative model, in a realistic environment, the generated energy deposits are processed by several reconstruction steps.
We investigate the quality of generated showers after processing with state-of-the-art reconstruction algorithms. \footnote{Different from e.g. Ref.~\cite{Sofia}, the goal here is not to replace reconstruction algorithms with another network, but to probe whether ``seamless'' integration of showers generated using a generative model into standard workflows would be possible.
}

The remainder of this paper is organised as follows: in Section~\ref{sec:data} we introduce the concrete
problem, the training data, and the particle flow reconstruction algorithm, 
followed by the considered network architectures in Section~\ref{sec:models}.
Section~\ref{sec:results} provides a quantitative evaluation of generated showers and Section~\ref{sec:conclusions}
gives conclusions and an outlook.


\section{Data and Reconstruction}
\label{sec:data}

We first introduce the dataset (Sec.~\ref{sec:dataset}) of hadron showers generated for this study.
As generated showers are compared both using the raw distributions as well as considering  the output of standard particle-flow-based reconstruction algorithms, these are discussed in Section~\ref{sec:reco}.

\subsection{Dataset}
\label{sec:dataset}

The International Large Detector (ILD)~\cite{ILD-IDR} is a proposed next generation particle detector at the International Linear Collider (ILC). It features highly granular sampling calorimeters optimized for the use of the particle flow reconstruction scheme. 

\begin{figure*}[h]
    \centering
    \includegraphics[width=0.65\textwidth]{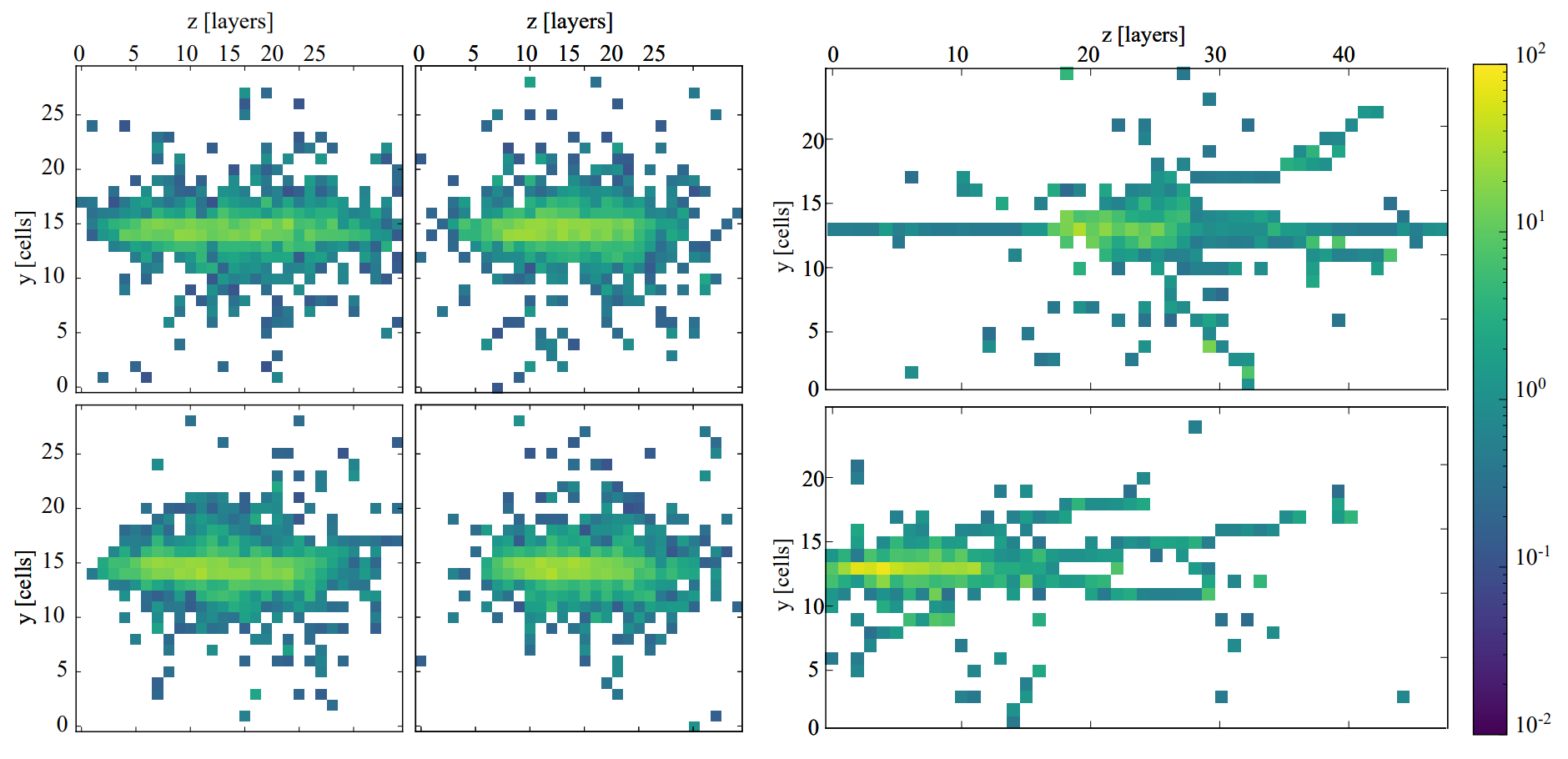}
    \includegraphics[width=0.33\textwidth]{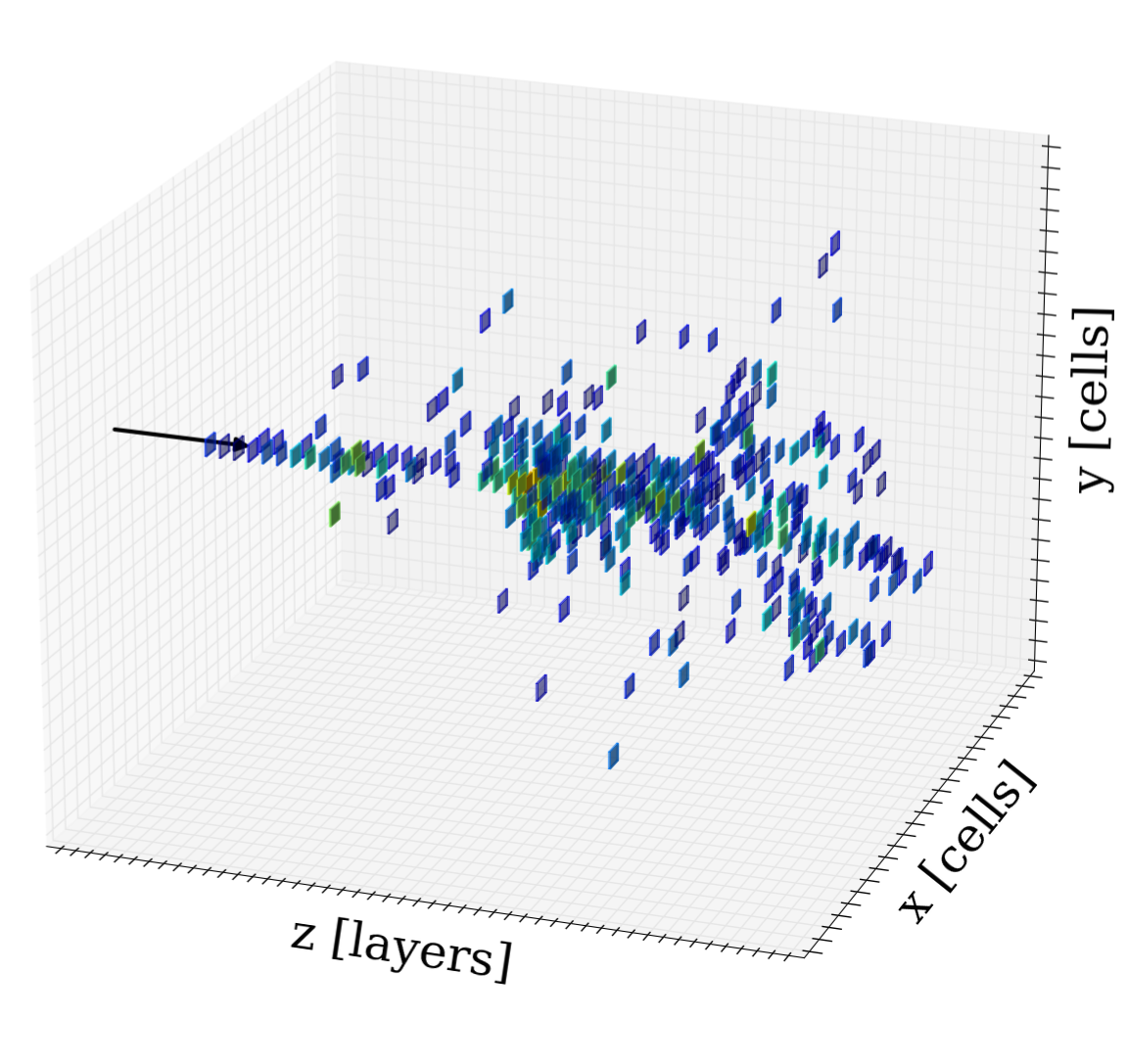}
    \caption{Left: four example images of photon showers. Center: two examples of pion showers. Right: 3D image of a pion shower. On average the pion showers have significantly more variance in their shapes.}
    \label{fig:shower_examples}
\end{figure*}

Particle showers occurring in these calorimeters are the simulation targets. However, unlike in previous work, which focused on photon showers in the SiW ECal~\cite{Buhmann:2020pmy}, we now consider showers initiated by positively charged pions in the highly-granular Analogue Hadron Calorimeter (AHCal). It consists of 48 layers with stainless steel absorber plates and scintillator tiles of 3$\times$3~$\textrm{cm}^2$ individually read out by Silicon PhotoMultipliers (SiPMs) as active material. The AHCal concept has been developed by the CALICE Collaboration~\cite{AHCAL_hardware}, and extensive beam tests~\cite{AHCAL_testbeam} have allowed a thorough validation of the simulation with real data. For this study, the ECal part of the calorimeter system, lying in front of the AHCal, is removed in order to avoid the additional complexity of handling different detector geometries and materials.
While the shape and behavior of photon showers is almost exclusively governed by electromagnetic interactions, charged pion showers also include hadronic interactions. This results in a much greater variety of shower topologies, as is illustrated in Fig.~\ref{fig:shower_examples}. The larger variety presents an increased challenge for generative networks.

ILD uses the iLCSoft~\cite{ilcsoft} software ecosystem for detector simulation, reconstruction and analysis. For the full simulation with \textsc{Geant4}~\cite{g4}, a detailed and realistic detector model implemented in \textsc{DD4hep}~\cite{dd4hep} is used. The training data consists of pion showers in the AHCal, which are simulated with \textsc{Geant4} version 10.4\footnote{using the \textrm{QGSP\_\,BERT} physics list} and \textsc{DD4hep} version 1.11. The iLCSoft ecosystem is fully containerized and simulation jobs are deployed into a \textsc{Kubernetes}~\cite{k8s} cluster hosted at DESY.

A virtual \textit{particle gun}
is placed at $(x',y',z') = (3\;\mathrm{cm}, 
100\;\mathrm{cm},100\;\mathrm{cm})$, shooting charged pions pointing along the $y'$-axis. 
Together with a strong axial magnetic field of 3.5 T, this leads to a roughly perpendicular incident angle onto the AHCal which is 106\;cm away from the pion gun. 
The \textit{primed} coordinate system of ILD is oriented such that the $z'$-coordinate points along the beam axis and the $y'$-axis points horizontally upwards. For the resulting training data, a coordinate system with the z-axis pointing along the depth of the calorimeter is used.
While the ECal is removed for the current study, the tracking system, which has a very low thickness in terms of radiation and interaction lengths, is not removed.
Combined simulation of the ECal and HCal 
will be considered in future work.



Our training data set consists of 500k charged pion showers with incident particle energies uniformly distributed between 10 and 100~GeV, and --- for each fixed incident energy --- a constant impact position and angle.\footnote{A subset of 5k example showers taken from the training set is available at \url{https://doi.org/10.5281/zenodo.5529677}.}
We project these calorimeter hits into a regular grid of $x \times y \times z = 25\times25\times48$ pixels.
Here, the $z$-axis is parallel to the particle trajectory and the 48 layers of the tensor correspond to the 48 layers of the calorimeter.
The grid is centered such that high-energy pions arrive at $(x,y,z)=(12,12,0)$, whereas
lower energetic particles (i.e. 10--20~GeV) enter at slightly shifted positions due to the magnetic field.
The size of the grid in the transverse direction was chosen as a trade-off between containing most of the deposited energy and keeping the image size (sparsity) from being too high (low).
For a 40~GeV pion, on average 96\% of the deposited energy is captured by the choice of $25\times25$.
We further correct for any artifacts caused by the slightly irregular calorimeter structure such that each cell in this grid corresponds to exactly one sensor.
In order to remove particles that pass through the detector without showering, we reject any shower with less then 70 hits above 0.25~MeV.
This requirement removes approximately $1\%$ of events.
In addition to the training set, we generate an independent test set to compare with the showers produced by the generative models. This consists of 49k showers with uniformly distributed particle energies. Additionally, for each particle energy in steps of 10 GeV from 20 to 90 GeV, smaller single-energy samples of 8k showers are produced.


\subsection{Particle Flow Reconstruction}
\label{sec:reco}

Typical calorimeters
at future $e^{+}e^{-}$ colliders are optimized for particle flow reconstruction. 
This means exploiting the high granularity of calorimeters combined with measurements from tracking detectors to reconstruct all individual particles created in an event.
As this will determine the physics performance that can ultimately be achieved, the quality of generated showers needs to be evaluated after passing through such a reconstruction algorithm.
In the following, we consider the state-of-the-art pattern recognition particle flow algorithm \textsc{PandoraPFA}~\cite{pandora_PFO}, as used by ILD.

The detailed full simulation with \textsc{Geant4} produces hit objects which consist of energy deposits in individual calorimeter cells as well as their exact positions in space. 
In a digitization step~\cite{ILD-IDR}, all effects such as 
readout electronics,
light yield of the scintillator tiles, and the statistical effects of the number of  SiPM pixels
are taken into account and applied to these simulated hits.
This is followed by a two step calibration procedure:
\begin{enumerate}
    \item The deposited energy is normalized to the most probable energy deposited by a minimum ionizing particle (MIP).
    \item This energy, in units of MIP, is converted into a total energy in GeV, such that the sum of all hit energies corresponds to the incident particle's energy. 
\end{enumerate}
After digitization and calibration, \textsc{PandoraPFA} is run to cluster the digitized calorimeter hits by iteratively applying a number of sophisticated clustering algorithms, using information of reconstructed charged particle tracks where applicable. The output of \textsc{PandoraPFA} is then a list of reconstructed particles, typically referred to as Particle Flow Objects (PFOs), which contain important information such as four-momentum, energy and particle ID. This list of PFOs is then directly used in subsequent analysis steps.


\section{Generative Models}
\label{sec:models}

We investigate the use of two distinct generative network architectures. The first setup is a rather lightweight Wasserstein GAN (Sec.~\ref{susec:wgan}) and the second is a significantly more complex Bounded Information Bottleneck AutoEncoder (Sec.~\ref{susec:bibae}). \footnote{The code for both the WGAN and BIB-AE including the hyperparamter settings used for training are available on \url{https://github.com/FLC-QU-hep/neurIPS2021_hadron}.}

\subsection{{Wasserstein GAN}}
\label{susec:wgan}

\begin{figure*}[h]
    \centering
    \includegraphics[width=0.6\textwidth]{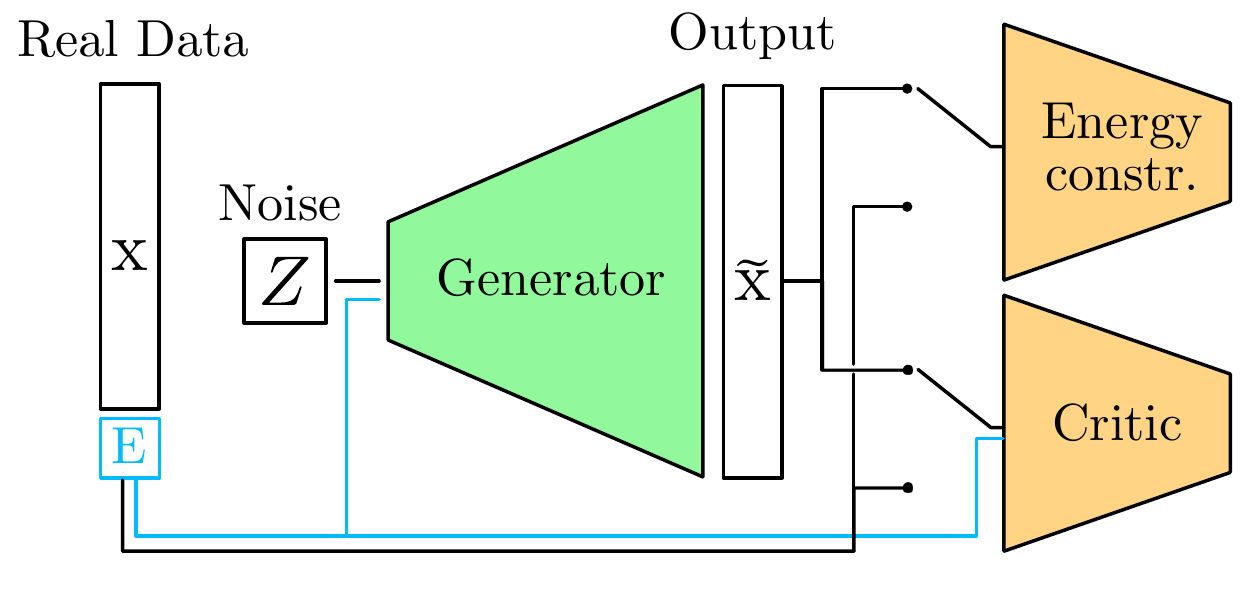}
    \caption{Schematic Illustration of our WGAN setup.}
    \label{fig:wgan_scheme}
\end{figure*}

The Wasserstein Generative Adversarial Network (WGAN) uses the Wasserstein-1 distance, also known as the earth mover's distance, as a loss function 
for better convergence and stability compared to classical GAN training~\cite{WGAN,WGAN2}. 
This distance evaluates the dissimilarity between two multi-dimensional distributions and informally gives the cost expectation for moving a mass of probability along optimal transportation paths~\cite{optimal_transport}.  Using the Kantorovich-Rubinstein duality, the Wasserstein loss can be expressed as

\begin{equation}
    \label{eqn:wloss}
    L = \textrm{sup}_{f\in \textrm{Lip}_1}\{\mathbb{E}[f(x)] - \mathbb{E}[f(\tilde{x})]\}.
\end{equation}

The supremum is over all 1-Lipschitz functions $f$, and is approximated by a discriminator network $D$ during the adversarial training. This discriminator is called the \textit{critic} since it is trained to estimate the Wasserstein distance between real and generated images. 

Furthermore, we need to ensure that the generated showers accurately resemble real showers of the requested energy $E$. This is achieved by parametrizing the generator and critic networks as functions of $E$ and by adding a constrainer~\cite{ErdmannWGAN2} network which returns the energy of a given shower.
It is trained prior to the WGAN in a fully supervised fashion, using only \textsc{Geant4} showers. 
The 100~million constrainer weights are then frozen during WGAN training. 

The WGAN  presented here is similar to the one used for photon shower generation in Ref.~\cite{Buhmann:2020pmy}, however two architectural changes are applied in order to improve its generative performance for hadronic showers.
\begin{itemize}
    \item The convolutional layers previously used in the critic network are replaced by 3D-residual blocks~\cite{3DResNet}. This change improves the expressiveness of the critic and increases the associated number of trainable parameters approximately four-fold to 4~million.
    \item A fully-connected network is employed instead of the convolutional layers previously used in the energy constrainer, as the shower patterns after preprocessing explicitly break translational symmetry.
\end{itemize}

The WGAN is trained for a total of 207k weight updates of the generator, which corresponds to 82 epochs. The \textsc{Adam} \cite{adam} optimizer is used with an initial learning rate of $10^{-4}$ ($10^{-5}$) for the generator (critic) networks. 



\subsection{Bounded Information Bottleneck Autoencoder}
\label{susec:bibae}

\begin{figure*}[h]
    \centering
    \includegraphics[width=0.97\textwidth]{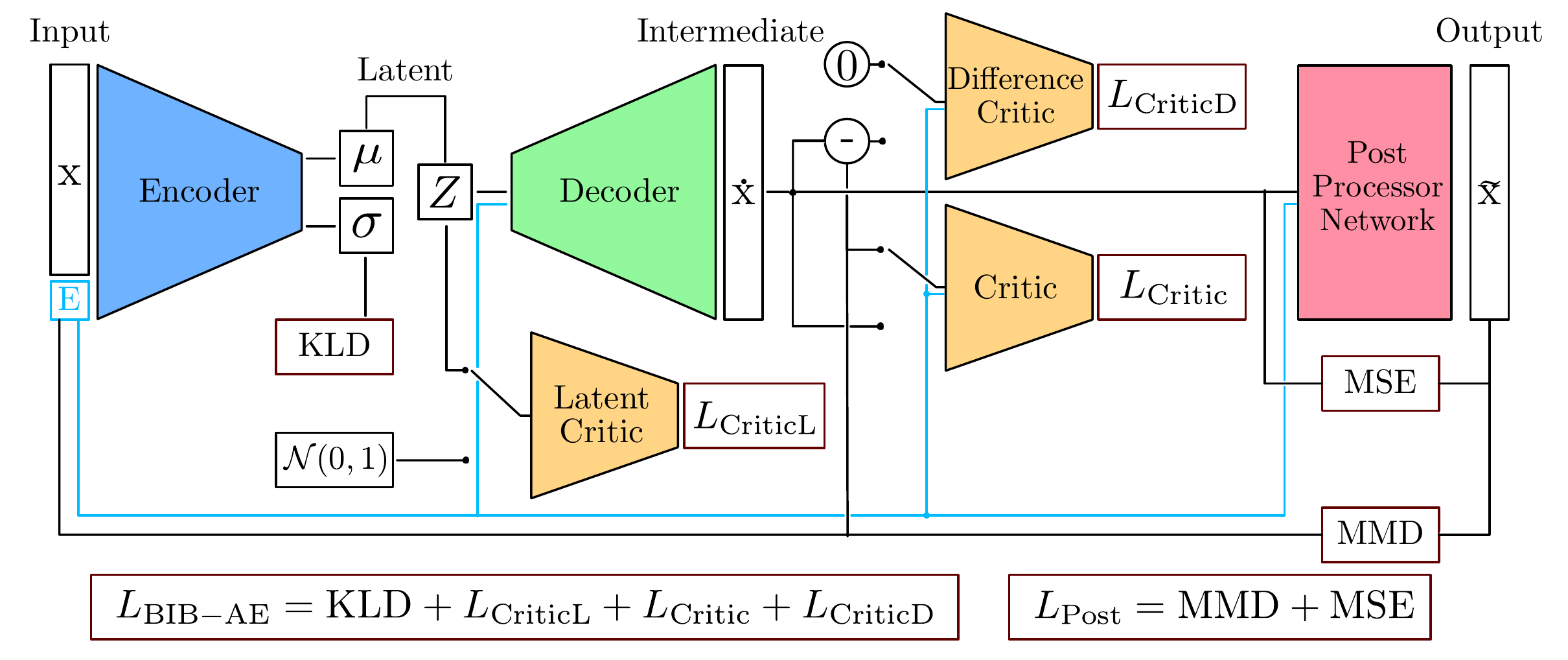}
    \caption{Schematic illustration of our BIBAE setup including the Post Processor Network in the final step.}
    \label{fig:BIBAE_scheme}
\end{figure*}




The Bounded Information Bottleneck Autoencoder (BIB-AE) unifies the most commonly used generative networks into an overarching theoretical framework~\cite{BIB-AE}. For this reason, many elements normally associated with these models can be found in the BIB-AE. Fundamentally speaking, it is an Autoencoder that maps from data-space to a lower-dimensional latent space and then back to  data-space. 

The BIB-AE presented here is a significant extension of the setup used for photon shower generation in Ref.~\cite{Buhmann:2020pmy}. 
We therefore first describe the baseline model followed by a detailed discussion of the extensions.
The baseline BIB-AE consists of two major components: The encoder/decoder chain that maps data to the latent space and back, and a series of auxiliary networks and loss terms used to train the main encoder and decoder. These auxiliary parts are: 
\begin{itemize}
    \item A dual purpose, Wasserstein-GAN like critic ($C$) that simultaneously judges whether the reconstructed output looks realistic and compares the output and input image to facilitate reconstruction. 
    \item A Kullback–Leibler divergence (KLD) that regularizes the latent space to ensure it has a shape close to a normal distribution.
    \item A second Wasserstein-GAN like critic ($C_L$) trained to differentiate between latent space and a normal Gaussian. 
    \item A Maximum Mean Discrepancy (MMD)~\cite{MMD_base} term comparing the latent space to a normal distribution to further regularize it. 
\end{itemize}
The contributions of these components are combined into a total loss term, each weighted by its own tuneable hyperparameter $\beta_{\alpha}$,

\begin{eqnarray}
    L_{\textrm{BIB-AE}} = & - \beta_{C_L} \cdot \mathbb{E}_{x\sim p_{data(x)}}[C_{L}(E(x))] \\
    & - \beta_{C} \cdot \mathbb{E}_{x\sim p_{data(x)}}[C(D(E(x)))] \\
    & + \beta_{\textrm{KLD}} \cdot \textrm{KLD}(E(x)) \\
    & + \beta_{\textrm{MMD}} \cdot \textrm{MMD}(E(x),\mathcal{N}(0,1)).
\end{eqnarray}

Both the BIB-AE network used in this work and the one previously used in~\cite{Buhmann:2020pmy} make use of a dedicated Post Processing Network to fine tune certain shower features. However the specific training process has been significantly modified.

\paragraph*{BIB-AE with KDE latent space sampling}
The underlying idea of generative Autoencoders is to map the data space to a well known latent space and back from this latent space to the data. When designing this latent space, one needs to strike a balance between regularization and expressiveness. A perfectly regularized latent space cannot contain information, while a very expressive latent space is difficult to sample from during generation. The standard approach in a VAE is to have a regularization loss term that is balanced with the reconstruction loss. This ensures that the latent space is both regularized and expressive. However for our BIB-AE setup the adversarial reconstruction loss makes this balance highly non-trivial. Instead we found larger success using a Buffer-VAE \cite{Buffer_VAE} inspired approach.

After training the main BIB-AE model, the encoder is used to translate the training data set into latent space samples. This latent space data provides a good description of the underlying latent space distribution. Our goal is now to draw new samples form this latent distribution. These samples can then, in turn, be passed to the BIB-AE decoder to generate new shower samples. We found several viable options to perform this latent sampling. Both a GAN and a Normalizing Flow are capable of reproducing the latent space very closely, however training these setups correctly is once again non-trivial. 
For this reason a Kernel Density Estimator (KDE)~\cite{KDE} is fitted to the latent space. 
As the latent space distribution depends on the energy, energy labels are included in the fit.
During generation, we sample both the particle energy label and the input latent noise from this KDE. In order to generate specific energy labels we make use of a rejection sampling method. 
A more in-depth discussion on the latent space sampling for the BIB-AE is provided in Ref.~\cite{decoding_photons}.

\paragraph*{{Minibatch discrimination}}
A major concern for physics applications of generative models is ensuring that the composition and overall properties of the generated data match that of the training set. 
Minibatch discrimination~\cite{improved_techniques_for_training} is a vital tool in this effort. The underlying idea is to give the discriminator network information about a whole batch of data, in addition to the information about the individual samples in that batch. This allows the discriminator to more easily spot outliers or mode collapse.   

Minibatch discrimination is implemented by first calculating the sum and standard deviation of each discriminator input sample. We then define a difference matrix between all of these sums and do the same for the standard deviations. These matrices are subsequently passed through an embedding network, in essence allowing the discriminator to learn the most important features of this batch-information. Finally, the outputs of this embedding are aggregated and passed to the fully connected main section of the discriminator. 
The above operations are carried out both for the un-scaled and log-scaled discriminator inputs.

\paragraph*{Dual and Resetting Critics}
One noticeable property of the pion shower data set is its sparsity. Despite having 30k possible pixel positions, the average number of active pixels above the MIP threshold (see sec~\ref{sec:results_genlevel}) is only 400, or about 1\%. This means, that over the course of the adversarial training, the critic network can become blind to certain pixel positions, which in turn gives rise to artifacts at these positions. One way to remedy this is to reinitialize the critic network. As this reinitialized network has no training history, it will easily spot these artifacts and force the decoder to correct them. However this has the downside that the critic never fully converges, making it harder to learn more subtle features. Therefore we replace every critic used in the BIB-AE with a set of two networks with identical architectures. One of these network is trained continuously, while the other has its weights and optimizer reset after each epoch.

\paragraph*{{Improved Post Processing}}
Much like the BIB-AE setup for photon showers, the pion setup also uses a dedicated Post Processor network. The training procedure has, however, been heavily modified. The first difference is that, while the photon Post Processor was trained parallel to the main BIB-AE, we now train it on a frozen version of the BIB-AE networks, after they have already finished training. While this does mean that the main BIB-AE model can no longer improve during the Post Processor training, it significantly stabilizes the Post Processor training and allows the network to be easily trained to convergence.   

Furthermore, the loss function of the Post Processor was also refined. In addition to the Mean Squared Error (MSE) and Sorted Kernel Maximum Mean Discrepancy (SK-MMD) described in~\cite{Buhmann:2020pmy} we implemented a series of auxiliary loss terms. 
\begin{itemize}
    \item The first SK-MMD is complemented with a second SK-MMD term with a larger kernel size. This allows for a larger coverage of the cell energy spectrum. 
    \item A Mean Squared Error/Mean Absolute Error loss between the sorted hit-energy values of the original input and processed shower. On the one hand this helps the Post Processor to better learn the energy spectrum, whilst on the other hand it also greatly improves the energy conditioning of the full setup.
    \item A loss term comparing the average shower image of an original input batch with the average of a processed shower batch. Previous iterations of the Post Processor were prone to adding artifacts to the shower, which were manifested in some pixels being significantly more likely to be active than others. These artifacts cannot be seen directly form individual images, but only in the averages over many showers. This loss term strongly reduces these artifacts.
\end{itemize}
The BIB-AE was trained for a total of 37 epochs. Using that model as a basis, the Post Processor was trained for 105 epochs. All networks used the \textsc{Adam}~\cite{adam} optimizer with exponential learning-rate decay. During training a threshold function was used to map every value below 10\% of a MIP to 0.  


In the following Section, we will observe the performance of the improved WGAN and BIB-AE networks for shower generation.
Due to the large computational cost of training, a detailed ablation study of the specific gains afforded by each modification was not possible.
However, without these improvements, no adequate description of hadronic shower distributions was possible.

\section{Results}
\label{sec:results}

In standard applications of generative models, the quality and fidelity of  individual generated images are often the most important metrics. 
However, when applied to physics simulation, the statistical distributions of specific quantities over many generated images become increasingly relevant.
This means we not only have to ensure that the individual showers look realistic, but also that the overall statistical properties of a set of showers agree with the training data.

In the following, we consider two different representation levels for comparisons between the output of generative models and the ground truth.
The first one is termed \textit{generator level}. 
As the name implies, at this stage, the direct output of simulation tools is compared, yielding insight into how well a specific model captures the training data. 
This is the default level used e.g. in Ref.~\cite{Buhmann:2020pmy} as well as other studies of generative models.


In practise, however, most physics analyses do not look directly at the raw measured or simulated data. Instead, a series of pattern recognition and data extraction algorithms are first applied, in a process referred to as reconstruction (see Sec.~\ref{sec:reco}). 
The result of this processing, the so-called \textit{reconstruction level}, constitutes the second data representation used for comparisons here. 
Considering a realistic reconstruction scenario is a major improvement of this work over previous results, as it allows us to judge which \textit{relevant} properties of the data a generative model learns.

Generator level results are presented in Section~\ref{sec:results_genlevel}, followed by reconstruction level results in Section~\ref{sec:results_recolevel}, and 
studies of the inference-time  in Section~\ref{sec:results_time}.

\subsection{Generator Level}
\label{sec:results_genlevel}

We first compare the output of the generative models with that of \textsc{Geant4} without additional reconstruction steps. 
The only additional processing performed is 
the removal of all cells with a pixel energy below half the energy deposition of a minimal ionizing particle, at 0.25 MeV.
In a real calorimeter, this selection removes noise produced by the detector components and any simulation therefore has to employ the same cutoff.

\begin{figure*}[tbh]
    \centering
    \includegraphics[width=0.29\textwidth]{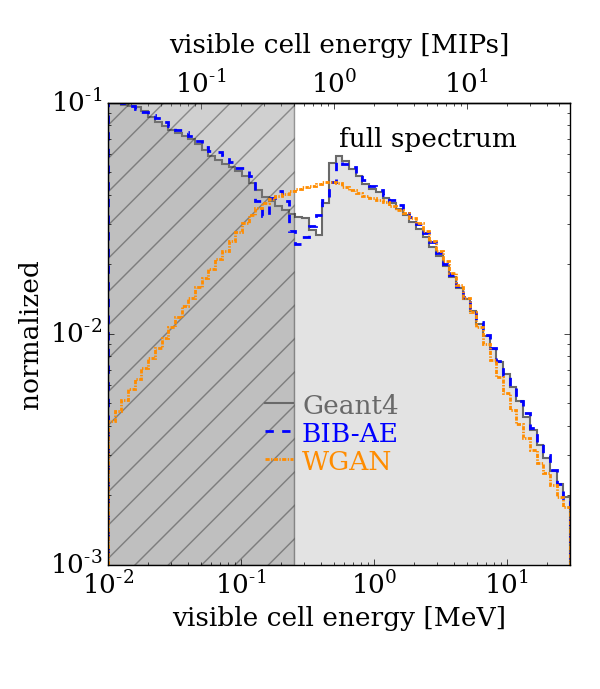}
    \includegraphics[width=0.29\textwidth]{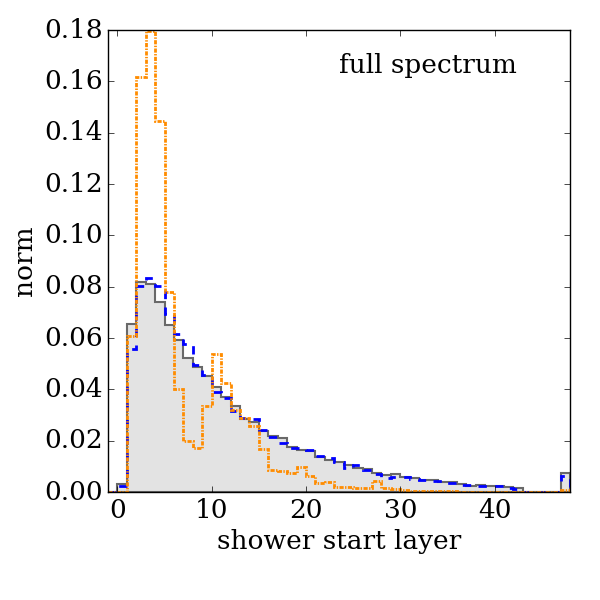}
    \includegraphics[width=0.29\textwidth]{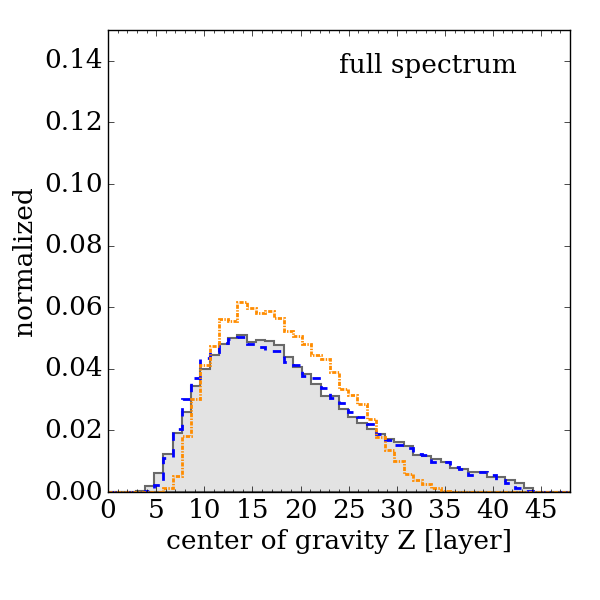}
    \includegraphics[width=0.29\textwidth]{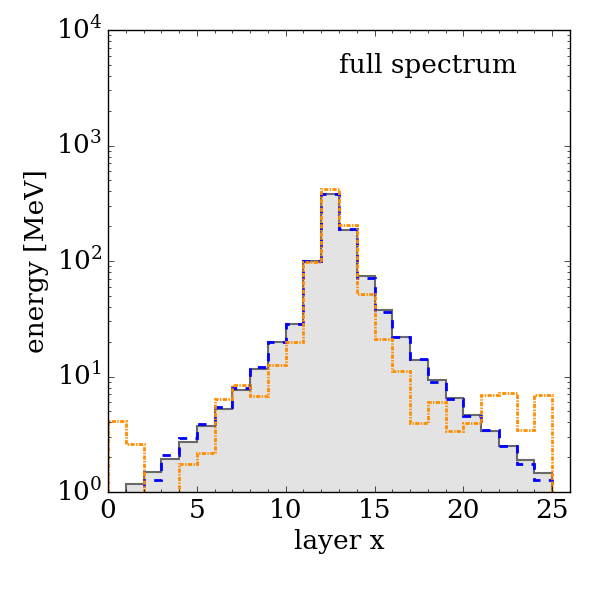}
    \includegraphics[width=0.29\textwidth]{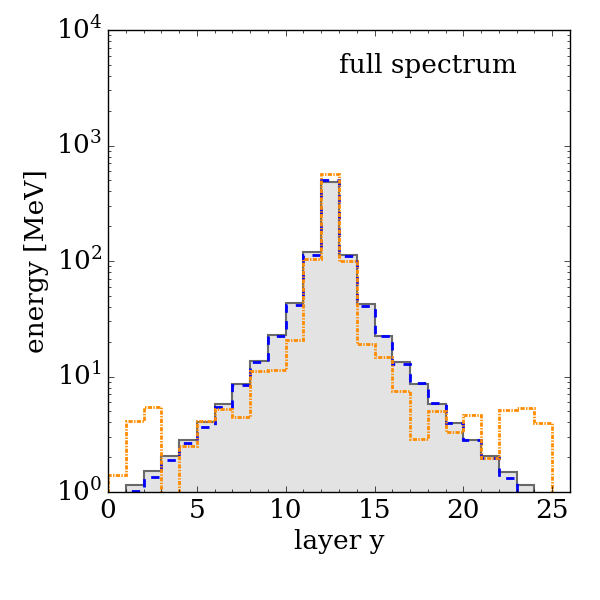}
    \includegraphics[width=0.29\textwidth]{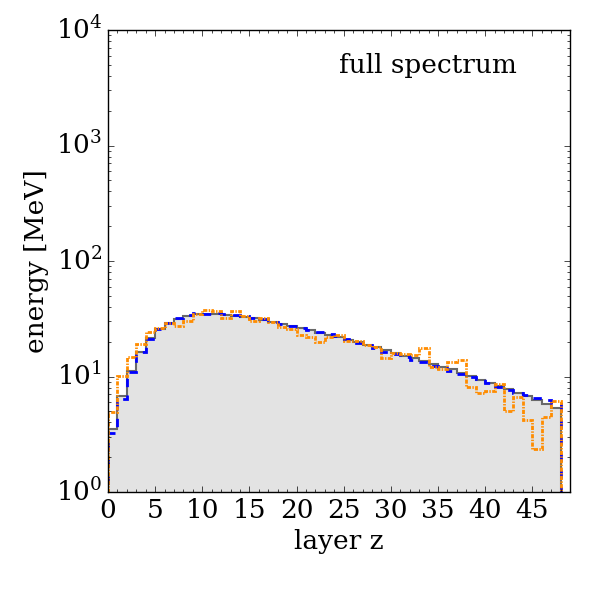}
    \caption{Differential distributions comparing physics quantities between \textsc{Geant4} and the different generative models for charged pion showers.
    In the top left, the energy per-cell is given in MeV on the bottom axis, and in multiples of the expected energy deposit of a minimum ionizing particle (MIP) on the top axis. The greyed out area indicates the 0.5 MIP cutoff. All plots are for showers generated with uniform energies in the 10-100 GeV range. 
    }
    \label{fig:gen_pion_shapedist}
\end{figure*}

Figure \ref{fig:gen_pion_shapedist} shows comparisons of physically relevant shower-shape observables.
Each plot uses the test-set generated using \textsc{Geant4} as the baseline (grey, filled) 
and compares it to the BIB-AE (blue, solid) and the WGAN (orange, dashed). The 10-100 GeV pion energy range of the dataset is used throughout.

In the top left plot we see the visible cell energy distribution. The hatched region of the plot indicates the cutoff at 0.5 MIP. As described above, any hit below this threshold is discarded for further comparisons. The main feature of interest is the peak located at 1.0 MIP. We can see that the BIB-AE setup, largely thanks to its Post Processor, accurately replicates this feature. 
The WGAN setup, on the other hand, seems unable to capture the peak. 
This is in line with other work \cite{Khattak:2021ndw, Buhmann:2020pmy} that has also demonstrated this difficulty in getting GANs to learn low energy features. 

The center top figure shows the calculated shower start position~\cite{shower_start}. Here we see a good agreement between \textsc{Geant4} and the BIB-AE, while the WGAN seems to produce a significant abundance of early starting showers. 
In the top right, we show the center of gravity  of the shower. This is equivalent to the first moment along the depth of the calorimeter, i.e the direction the particle is traveling in. Once again, the BIB-AE distribution overlaps almost perfectly with the \textsc{Geant4} one. The WGAN shows a significant overlap, but fails to correctly model the tail regions. 
The bottom left, center and right plots show the average energy profiles of the showers in the $x$, $y$, and $z$ direction respectively. For all 3 profiles, the BIB-AE showers line up with the \textsc{Geant4} ones, save for minor fluctuations in the tail regions. Specifically, the asymmetry in the $x$ profile caused by the magnetic field in the detector is also correctly modeled. 
The WGAN on the other hand exhibits significant artifacts.

\begin{figure*}[h]
    \centering
    \includegraphics[width=0.45\textwidth]{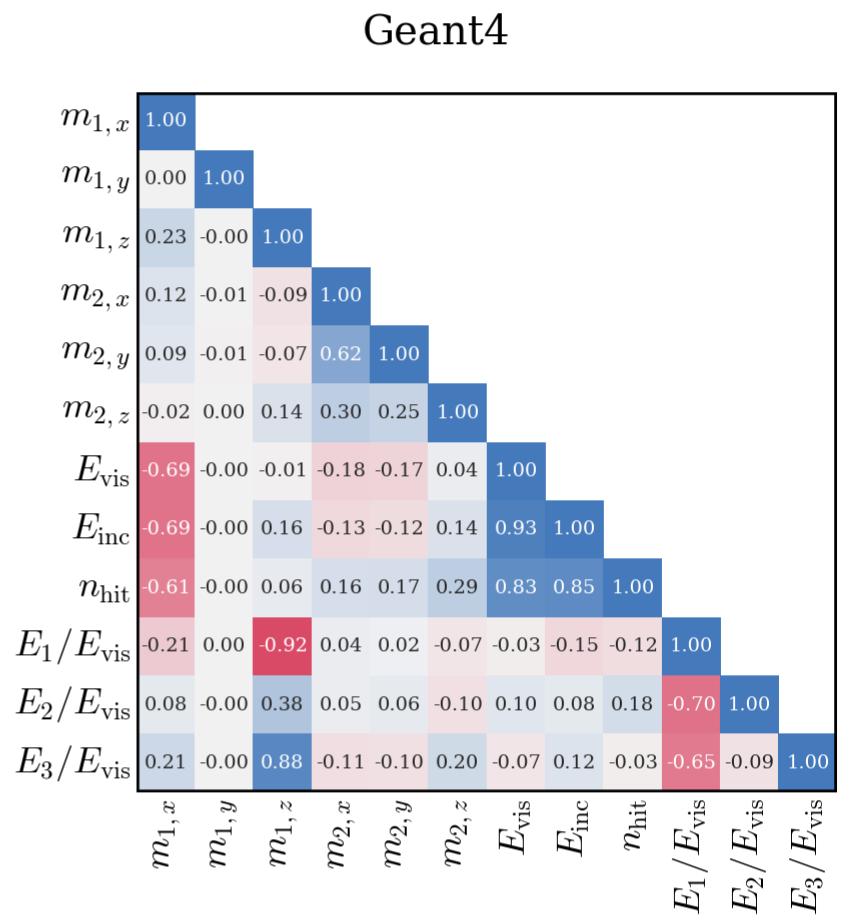}
    \includegraphics[width=0.45\textwidth]{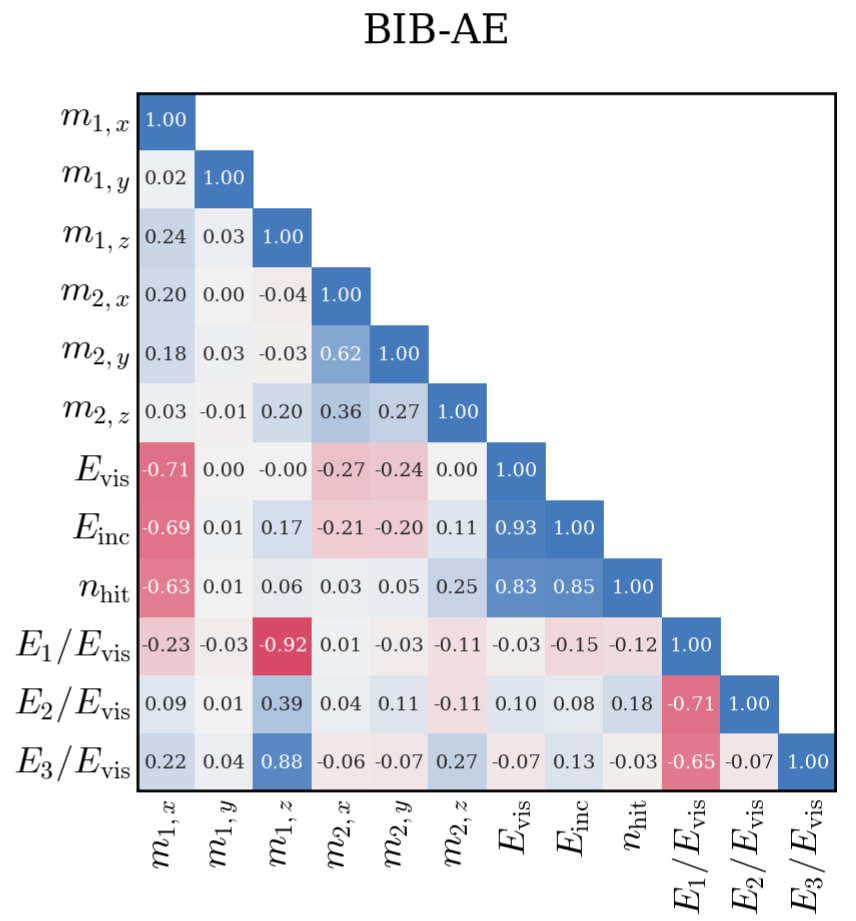}
    \includegraphics[width=0.45\textwidth]{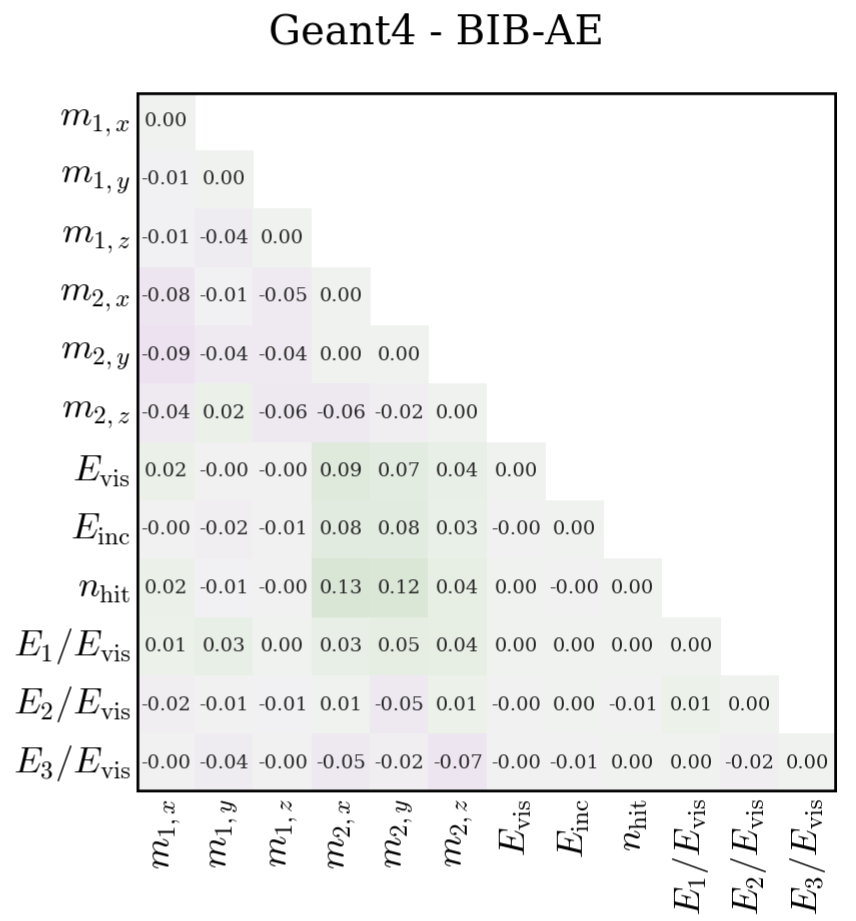}
    \includegraphics[width=0.45\textwidth]{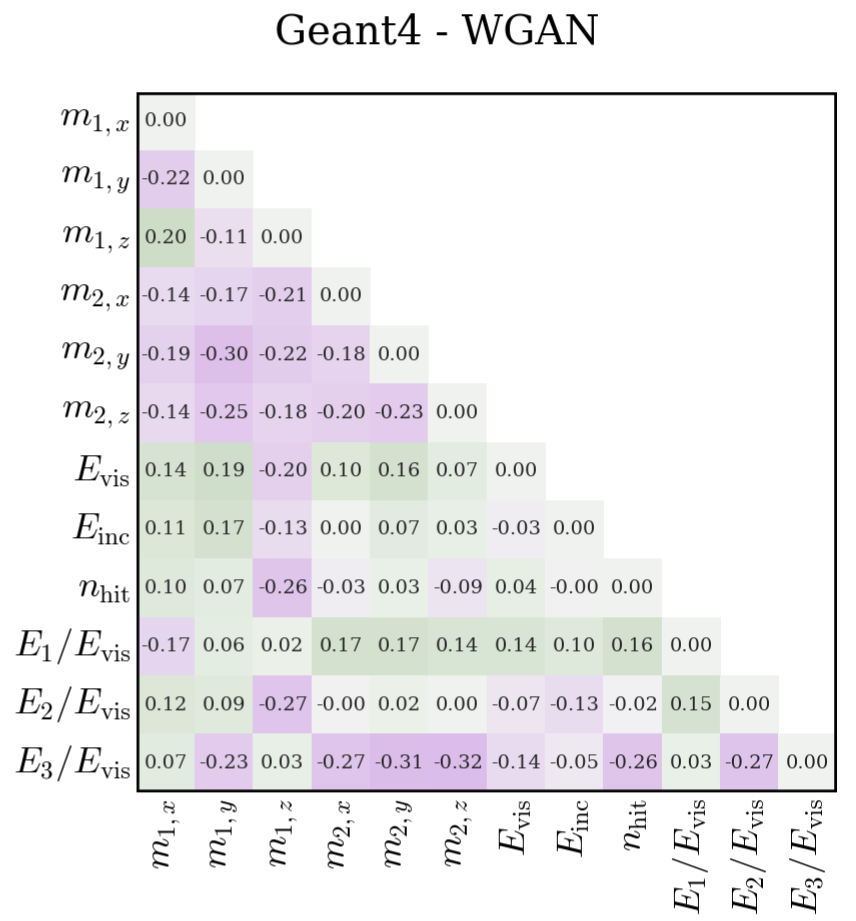}
    \caption{Top: Pearson correlation coefficients for selected physical parameters of showers simulated with \textsc{Geant4} (left) and generated with the BIB-AE (right). See the text for specific definitions of the individual parameters. Bottom: Differences between these correlation coefficients and \textsc{Geant4} for the BIB-AE (left) and WGAN (right). 
    }
    \label{fig:gen_pion_corr}
\end{figure*}

In addition to the marginal distributions of individual observables, a successful generative model also needs to learn higher-dimensional correlations. 
To this end, we explicitly verify the correlations between all pairs of considered observables.
The top row of Fig.~\ref{fig:gen_pion_corr} shows two correlation matrices, with each entry corresponding to the Pearson correlation coefficient between the following quantities: First moment in x, y, and z direction ($m_1$), second moment in x, y, and z direction ($m_2$), the total visible energy ($E_{\textrm{vis}}$), the incident particle energy ($E_{\textrm{inc}}$), the number of hits above the 0.5~MIP threshold ($n_{\textrm{hit}}$) and the energy fractions deposited in the first, second, and last third 
along the depth of the calorimeter ($E_i/E_{\textrm{vis}}$).
We perform this correlation calculation for \textsc{Geant4}, the BIB-AE and the WGAN. In order to best compare the correlations produced by the generative models and those of \textsc{Geant4}, the BIB-AE and WGAN matrices are subtracted from the \textsc{Geant4} matrix. These results are shown in the bottom of Fig.~\ref{fig:gen_pion_corr}. The closer to zero the differences are, the better the correlations are reproduced. We see that for the BIB-AE the largest deviation is around $0.13$, while most do not exceed $0.05$. This is indicative of the ability of the BIB-AE to reproduce the \textsc{Geant4} correlations. For the WGAN we see significantly larger deviations, up to $0.31$. This is in line with previous results showing the difficulties the WGAN has in learning properties such as the shower energy profiles.

\begin{figure*}[h]
    \centering
    \includegraphics[width=0.45\textwidth]{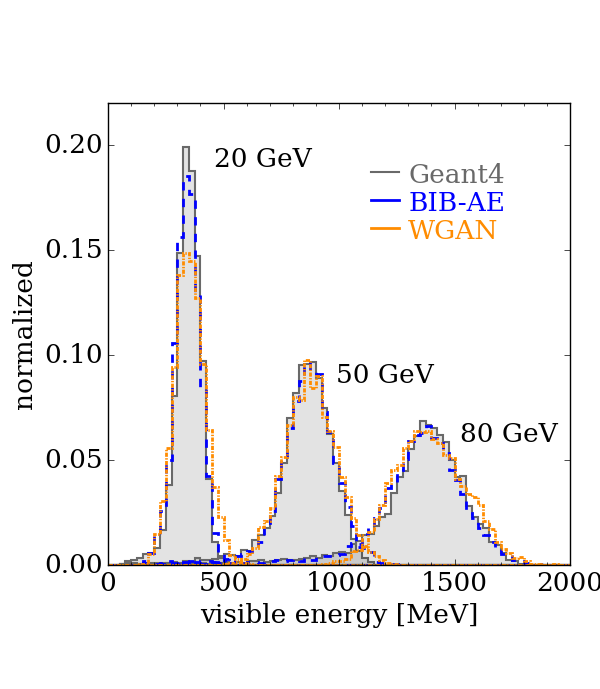}
    \includegraphics[width=0.45\textwidth]{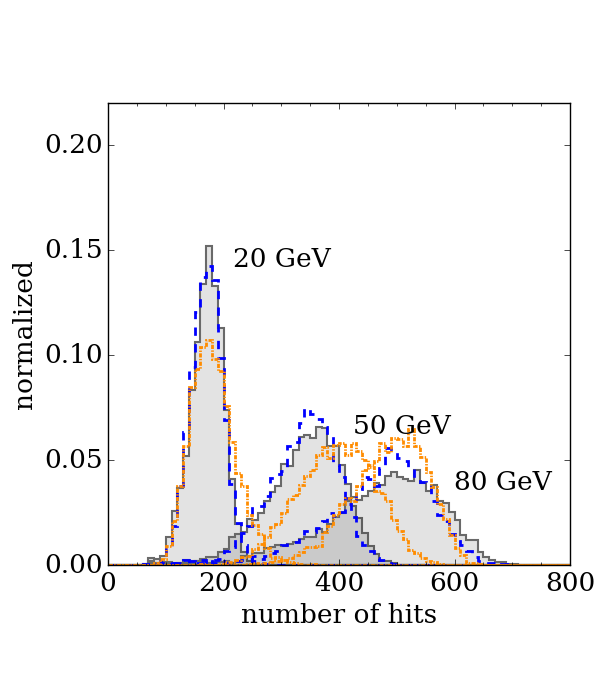}
    \caption{
    Comparison of the visible energy and number of active hits between \textsc{Geant4} and the different generative models for three selected pion energies.
    The number of hits in the right plot is calculated after applying a cutoff at 0.5 MIP.}
    \label{fig:gen_pion_conddist}
    \hspace{0.5cm}
\end{figure*}

The third set of comparison plots is presented in Fig.~\ref{fig:gen_pion_conddist}. The color coding remains the same as in the first set, however only discrete pion energies of 20, 50 and 80 GeV are used. This allows for explicit testing of whether the energy conditioning is learned correctly or not. 
The leftmost plot shows the total energy deposited in the active regions of the calorimeter.
Both the energy sums of the BIB-AE and the WGAN largely match those present in \textsc{Geant4}. Furthermore, while both networks seem to slightly mismodel the very sharp 20~GeV peak, they perform very well for the 50 and 80~GeV peaks, where they correctly model the means and widths of the peaks. 
The plot on the right shows the total number of pixels with values above the MIP threshold. One vital aspect of modeling this quantity is correctly capturing the visible cell energy spectrum around the cutoff (Fig.~\ref{fig:gen_pion_shapedist}, top left), as even small shifts in this region can have large impacts on how many points end up above or below the cutoff. This explains why the BIB-AE is very successful in reproducing the total number of hits, while the WGAN shows some significant deviation, especially for the 50~GeV distribution. 

\begin{figure*}[h]
    \centering
    \includegraphics[width=0.45\textwidth]{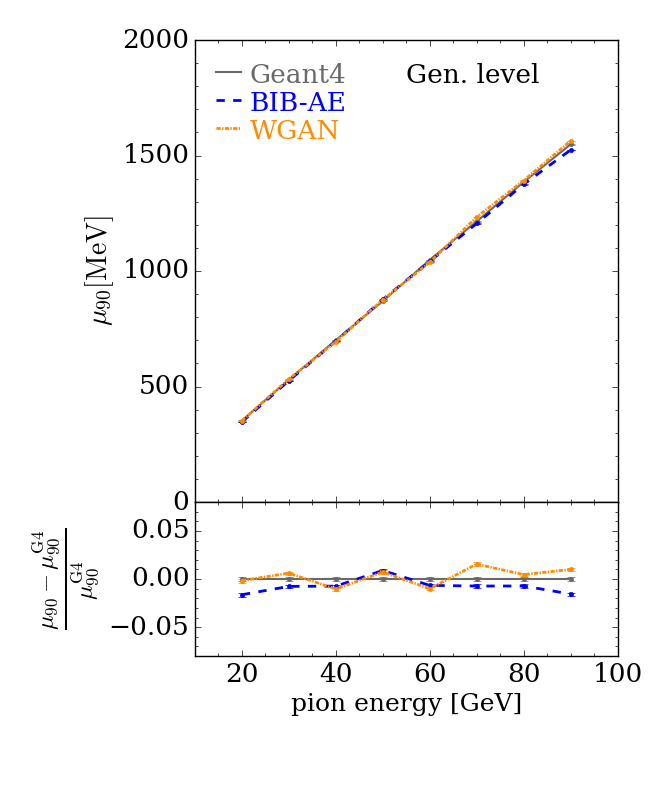}
    \includegraphics[width=0.45\textwidth]{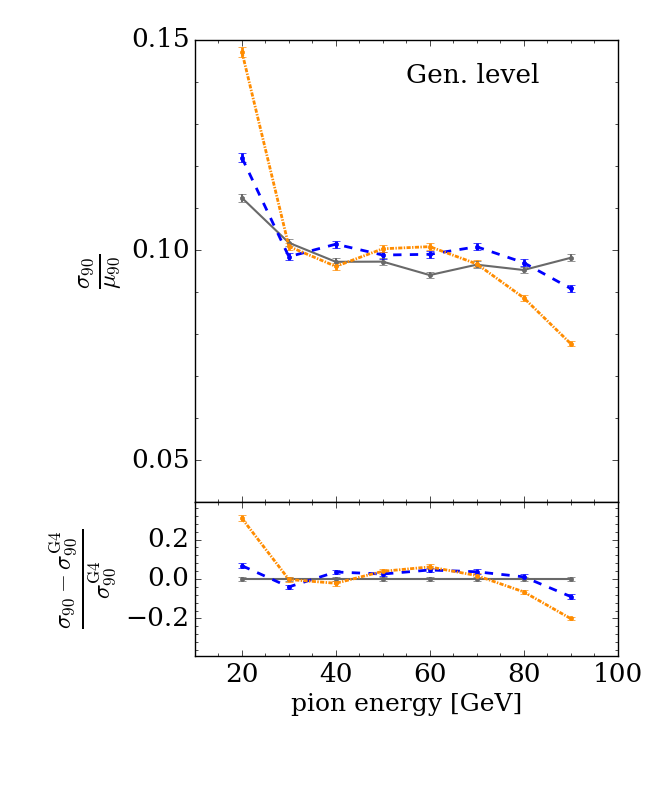}
    \caption{Mean ($\mu_{90}$, left) and relative width ($\sigma_{90}/\mu_{90}$, right) at generator level for pions with various incident energies. In order to avoid edge effects, the phase space boundary regions of 10 and 100 GeV are removed for the response and resolution studies. In the bottom panels, the relative offset of these quantities with respect to the Geant4 simulation is shown.
    }
    \label{fig:gen_pion_condlin}
\end{figure*}

Figure \ref{fig:gen_pion_condlin} shows a more in-depth look at the energy conditioning. For fixed energies between 20 and 90~GeV we produce a set of showers using both \textsc{Geant4} and the generative models. We then calculate the visible energy sums of these showers and determine the mean and root-mean-square of the 90\% core of these distributions, labeled $\mu_{90}$ and $\sigma_{90}$ respectively, for all energies. Finally we plot the means and relative widths as a function of the incoming particle energy. Note that 10 and 100~GeV were omitted from this study as these points lie right at the phase space boundaries. For energies above 40~GeV the resolution does not improve with energy due to leakage effects becoming important.
The leftmost curve shows that the position of the mean is especially well captured by the WGAN, with a maximum deviation of 2\%. 
The BIB-AE exhibits some larger discrepancies, up to 3\% in the high and low energy sections, but still provides a reasonable agreement. Note that a calibration factor has been applied to the WGAN-generated single-energy showers to improve the linearity. The relative width in the right plot is not modeled as well, exhibiting differences up to the 10\% level for the BIB-AE and up to the 30\% level for the WGAN in the edge regions. 

\subsection{Reconstruction Level}
\label{sec:results_recolevel}

To apply the \textsc{PandoraPFA} reconstruction software, the outputs of the WGAN and BIB-AE, which yield (batches of) $25\times25\times48$ \textsc{NumPy} \cite{harris2020array} arrays, need to be converted back  into actual cell positions and hit energies in the ILD calorimeter. 
These hits are then provided as input to \textsc{PandoraPFA} for reconstructing corresponding PFOs\footnote{As no track reconstruction is considered, \textsc{PandoraPFA} will reconstruct neutral PFOs using calorimeter information only.}.
The resulting PFOs are then compared to those created with the standard \textsc{Geant4} simulation-reconstruction chain. To ensure a consistent comparison, the \textsc{Geant4} data undergoes the same projection/conversion operation as the WGAN and BIB-AE.

\begin{figure*}[h]
    \centering
    \includegraphics[width=0.45\textwidth]{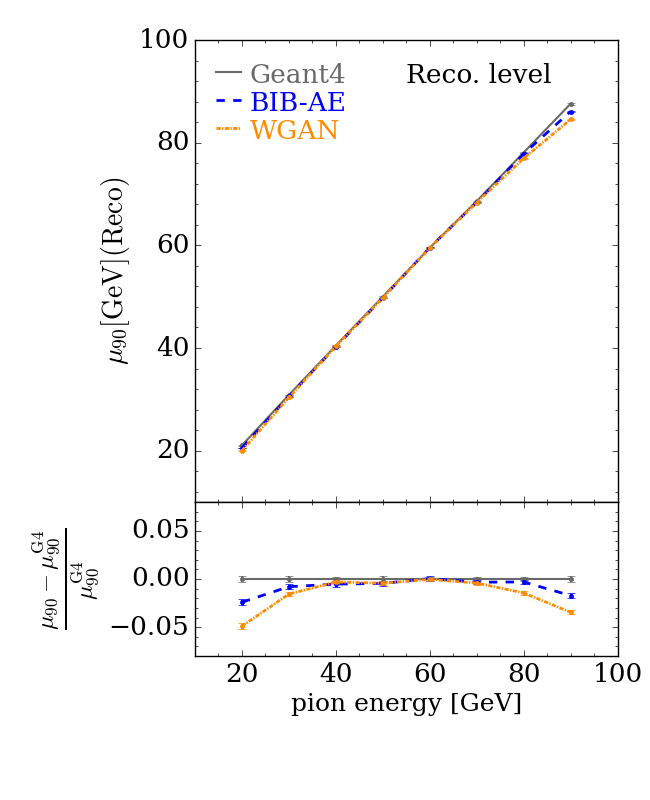}
    \includegraphics[width=0.45\textwidth]{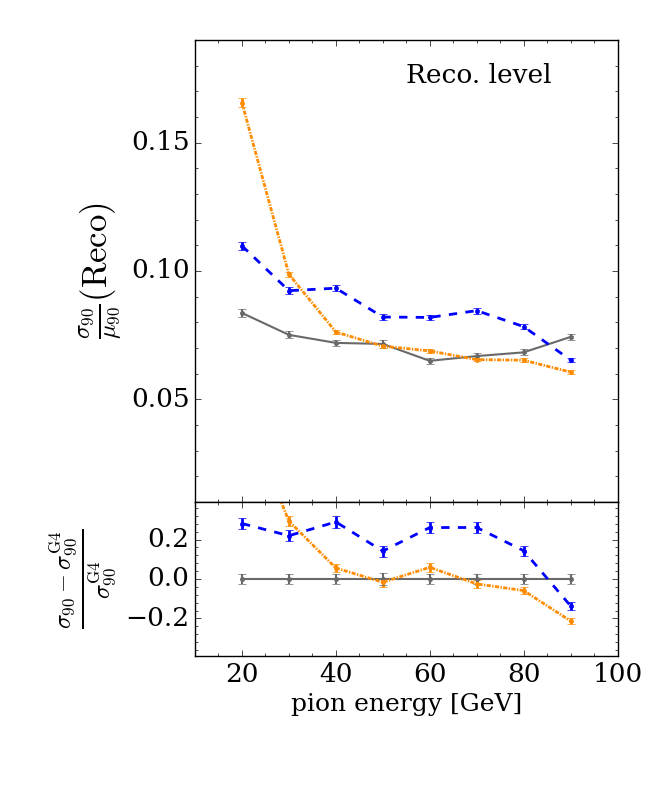}
    \caption{Mean ($\mu_{90}$, left) and relative width ($\sigma_{90}/\mu_{90}$, right) at reconstruction level for pions with various incident energies. In order to avoid edge effects, the phase space boundary regions of 10 and 100 GeV are removed for the response and resolution studies. In the bottom panels, the relative offset of these quantities with respect to the Geant4 simulation is shown.
    }
    \label{fig:reco_pion_condlin}
\end{figure*}

Figure~\ref{fig:reco_pion_condlin} shows the same quantities presented in Fig.~\ref{fig:gen_pion_condlin}, but now at the reconstruction level. 
The leftmost plot shows that the position of the mean is well captured in the middle range of energies by both the models. 
Likewise, both models display some larger discrepancies, up to 3-5\% in the high and low energy sections, but still have a reasonable agreement with \textsc{Geant4}. 
The relative width on the right plot shows a fairly good agreement for the WGAN for the middle incident energies. On the edge regions, however, up to 20\% differences for the BIB-AE and up to 40\% for the WGAN are present. It is worth noting that our models and \textsc{Geant4} have better relative width compared to the generator level as \textsc{PandoraPFA} uses a software compensation algorithm~\cite{pandora_sc} that improves the energy reconstruction of clusters by weighting hits depending on their hit energy density.

\subsection{Computing Times}
\label{sec:results_time}
The prime objective for using generative models in particle physics is to reduce the time and cost per simulated sample. To do so, we benchmark the per-shower generation time both on CPU and GPU hardware architectures. Fixed factors, such as initial sample generation and network training time, are not included in this accounting, as they are expected to be small compared to the overall number of showers to be generated.
Table~\ref{table:bench} shows the average time to generate a shower with an energy in the 10-100 GeV range using \textsc{Geant4}, the WGAN, and the BIB-AE. Both models offer significant speedups compared to classical generation methods. Furthermore, we also see trade-offs between these models. While the BIB-AE produces overall better quality showers than the WGAN, it also is one order of magnitude slower. 
\begin{table*}[h!]
\sisetup{
separate-uncertainty=true,
table-format=4.3(5)
}
\centering

\caption{Computational performance of WGAN and BIB-AE generators on a single core of an Intel\textsuperscript{\tiny\textregistered} Xeon\textsuperscript{\tiny\textregistered} CPU E5-2640 v4 (CPU) and NVIDIA\textsuperscript{\tiny\textregistered} A100 with 40~GB of memory (GPU) compared to \textsc{Geant4}. For the generative models, the best performing batch size is shown and given by the mean and standard deviation obtained for sets of 10000 showers.}\label{table:bench}
\vspace{15pt}
\begin{tabular}{ll|Sr}
\toprule
Hardware & Simulator &  {Time / Shower [ms]} & Speed-up \\\midrule
CPU & \textsc{Geant4} & 2684 \pm 125 & $\times 1$ \\
    & & & \\
    & WGAN & 47.923\pm0.089 & $\times 56$\\
    & BIB-AE & 350.824\pm0.574 & $\times 8$ \\
    & & &\\
GPU & WGAN & 0.264\pm0.002 & $\times 10167$ \\
    & BIB-AE & 2.051\pm0.005 & $\times 1309$ \\\bottomrule
\end{tabular}
\vspace{15pt}
\end{table*}

\section{Conclusions}
\label{sec:conclusions}

The strategy of using generative models to augment classical simulations has made rapid progress since its inception.
This paper advances the state-of-the-art in two regards: learning more complex hadronic showers in a high-granularity calorimeter, and considering the effects of particle-flow reconstruction algorithms.

We observe that the modified BIB-AE achieves excellent agreement with all physical observables at generator-level, both over the full spectrum and for specific incident particle energies.
The largest disagreement of $\approx 10\%$ is seen for the hit multiplicity distribution, while most other observables agree to the percent-level or better.
This is made possible by using a second-stage density estimator to sample from the learned latent space distributions, utilizing batch-level information, adding a resetting critic network, and a more convergent post-processing network.

Still on generator level, the performance of the much simpler WGAN architecture considered is clearly worse. For example, it does not learn the correct energy distribution around the MIP-value which leads to a mismodelling of the distribution of the number of hits. Similarly, the longitudinal shower profile shows unphysical structure.

However, considering the offset and width of reconstructed energies, the difference is much smaller. Both BIB-AE and WGAN achieve excellent linearity with the largest deviation of 5~\% observed at the boundaries of the considered energy range. For incident particle energies between 40 and 80~GeV the WGAN-generated showers also accurately track the width of the underlying \textsc{GEANT4} simulation.
Only the mean energy and width, as the most relevant quantities for physics analysis, were so far considered at the reconstruction level. Future work will extend the investigation also to other properties of the shower.

Finally, we again confirm the previously observed speed-up over the initial \textsc{Geant4} generation when sampling from generative models. The more complex BIB-AE architecture is slower than the simpler WGAN by approximately a factor of eight.
The main cause for this is that BIB-AE uses transpose convolutions to get to the full image size and then runs another set of convolutions at that full size, while the WGAN applies no additional convolutional layers once the full image size has been reached.
The maximum speed-up of four orders of magnitude is observed when comparing WGAN (GPU) to \textsc{GEANT4} executed on CPU. 

It is interesting to observe that while generative models overall offer a trade-off between precise simulation and resource consumption, this also holds true when comparing different generative architectures.
The successful and accurate simulation of hadronic showers is another major milestone towards application-ready generative models for highly-granular calorimeters.


\ack

This research was supported in part through the Maxwell computational resources operated at Deutsches Elektronen-Synchrotron DESY, Hamburg, Germany.
E. Buhmann and W. Korcari are funded by the German Federal Ministry of Science and Research (BMBF) via  \textit{Verbundprojekts 05H2018 - R\&D COMPUTING
(Pilot\-maß\-nah\-me ErUM-Data) Innovative Digitale
Technologien f\"ur die Erforschung von Universum und
Materie}. 
S. Diefenbacher is funded by the Deutsche Forschungsgemeinschaft (DFG, German Re\-search Foundation) under Germany’s Excellence Strategy – EXC 2121  ``Quantum Universe" – 390833306.  
E. Eren is funded through the Helmholtz Innovation Pool project ACCLAIM that provided a stimulating scientific environment for parts of the research done here. L. Rustige was supported  by  DESY and HamburgX grant LFF-HHX-03 to the Center for Data and Computing in Natural Sciences (CDCS) from the Hamburg Ministry of Science, Research, Equalities and Districts. This project has received funding from the European Union’s Horizon 2020 Research and Innovation programme under Grant Agreement No 101004761.

\vspace{1cm}

\section*{References}

\bibliographystyle{iopart-num-mod}
\bibliography{literature.bib}

\end{document}